\documentclass[aps,pra,reprint,showpacs,superscriptaddress]{revtex4-1}
\usepackage{amsmath,amssymb,mathrsfs,amsfonts}
\usepackage{epsfig}
\usepackage{bm}
\usepackage{color}
\usepackage[dvipdfm,
            pdfstartview=FitH,
            CJKbookmarks=true,
            bookmarksnumbered=true,
            bookmarksopen=true,
            linktocpage=true,
            colorlinks=true, % 注释掉此项则交叉引用为彩色边框(将colorlinks和pdfborder 同时注释掉)
            pdfborder=001,   % 注释掉此项则交叉引用为彩色边框
            citecolor=blue,  % for the color of citation links to the bibliography
            urlcolor=blue,   % for the color of URL links like web or mail addresses
            linkcolor=blue,  % for the color of internal links to sections, pages etc.
            anchorcolor=blue,
            ]{hyperref}      % hyperref 宏包通常要求放在导言区的最后!!!

\begin{document}
%\begin{CJK*}{GBK}{}
  \title{Quantum coherent dynamics in single molecule systems: generalized stochastic Liouville equation}
  \author{Xiangji Cai}
  \email{xiangjicai@foxmail.com}
  \affiliation{School of Physics, Shandong University, Jinan 250100, China}
  \affiliation{School of Science, Shandong Jianzhu University, Jinan 250101, China}

  \author{Yujun Zheng}
  \email{yzheng@sdu.edu.cn}
  \affiliation{School of Physics, Shandong University, Jinan 250100, China}

\begin{abstract}
  We propose the generalized stochastic Liouville equation to investigate
  the coherent dynamics in single molecule systems coupled to environments
  which exhibit both nonstationary and non-Markovian features.
  The generalized stochastic Liouville equation contains a generalized
  memory kernel associated with both the intrinsic system Hamiltonian and
  the non-Markovian features of the environmental noise, which returns to
  the well-known framework established by Kubo in the limit case that the
  environmental noise is stationary and memoryless.
  The coherence of the quantum system can be derived by means of the
  generalized stochastic Liouville equation in two separated averages
  with no need to consider the statistical characteristics of the environmental noise.
  We express the exact analytical expressions of the coherence of the single
  molecule systems induced by the nonstationary and non-Markovian RTN,
  and the analytical results of the system coherence are in well consistent with
  that derived by means of some other theoretical approaches.
\end{abstract}

% 03.65.Yz  Decoherence; open systems; quantum statistical methods
% 05.40.-a  Fluctuation phenomena, random processes, noise, and Brownian motion
% 02.50.-r  Probability theory, stochastic processes, and statistics
% 02.50.Cw  Probability theory
% 02.50.Ey  Stochastic processes
% 02.50.Fz  Stochastic analysis
% 05.10.Gg  Stochastic analysis methods (Fokker-Planck, Langevin, etc.)
\maketitle

\section{Introduction}

Coherence as an important nature of quantum world has been demonstrated
to be a useful resource which plays a crucial role in physical, chemical
and biological communities~\cite{Nature446.782,Nature463.644,
ProcNatlAcadSci.107.12766,NatPhys.7.448,Science340.1431,Science340.1448,RevModPhys.89.041003}.
Investigations on quantum coherent dynamics in single molecule have draw
much extensive attention due to our increasing capability to observe and
control quantum systems at the single-molecule level.
At such a scale, there are constantly revealing many new physical effects
and mechanisms where the decoherence caused by the environments strongly
influences the dynamical evolution of the single molecule~\cite{PhysRevLett.90.238305,
AnnuRevPhysChem.55.457,PhysRevX2.031012,RevModPhys.87.1153,RevModPhys.87.1183,PhysRep.831.1,PhysRep.838.1}.
Environmental effects on quantum systems can be modeled by random telegraph noise (RTN)
which plays an important role in some fundamental physical, chemical and biological processes,
such as, rate process in chemical reactions~\cite{RevModPhys.62.251}, optical trapping in
biological cells~\cite{Nature330.769,ProcNatlAcadSci.94.4853}, stochastic
resonance in excitable systems~\cite{RevModPhys.70.223,PhysRep.392.321},
and frequency modulation in quantum information science~\cite{RepProgPhys.80.056002}.
In addition, RTN as an important non-Gaussian noise has also been used to generate the low-frequency
$1/f^{\alpha}$ noise both theoretically and experimentally~\cite{RevModPhys.60.537,PhysRevLett.88.228304,PhysRevLett.92.117905,PhysRevLett.96.097009,PhysRevB74.024509,
PhysRevB79.125317,PhysRevB86.205404,PhysRevA85.052125,PhysRevA87.052328,RevModPhys.86.361,ApplPhysLett.110.081107,PhysRevA100.052104},
which are responsible for coherent dynamics in quantum solid-state nanodevices.

For a long time, the coherent dynamics of an open quantum system such as a
single molecule coupled to its environments is usually governed by a Lindblad
type master equation within the Markov approximation, which assumes that the
information flows unidirectionally from the system into the
environments~\cite{JMathPhys.17.821,CommunMathPhys.48.119}.
With the development of experimental technique, it has been observed accurately
that the dynamical evolution of open quantum systems is closely associated with
a backflow of information from the environments into the system.
For instance, the processes of electronic energy transfer in photosynthesis
and decoherence in quantum bits display strong non-Markovian behavior~\cite{Nature446.782,Nature463.644,ProcNatlAcadSci.107.12766,NewJPhys.10.113019,
JChemPhys.131.105106,NewJPhys.11.033003,JChemPhys.138.204309}.
In the recent two decades, the non-Markovian effect of the coherent dynamics in
open quantum systems has drawn increasing attention in a wide variety of fields
due to its key role in the community of physics, chemistry and biology
systems~\cite{NatPhys.7.931,NatChem.4.568,PhysRevLett.103.210401,PhysRevLett.105.050403,PhysRevLett.112.120404,
RepProgPhys.77.094001,RevModPhys.88.021002,RevModPhys.89.015001,JPhysChemLett.8.5390}.
Many excellent theoretical approaches have been proposed to investigate the non-Markovian dynamics of open quantum systems, such as, quantum state diffusion~\cite{PhysRevA58.1699,PhysRevA.60.91,PhysRevLett.82.1801}, projection operator~\cite{PhysRevA59.1633,PhysRevB70.045323,PhysRevE72.056106,PhysRevE73.016139,Entropy21.1040}, quantum jumps~\cite{PhysRevLett.100.180402,PhysRevA79.062112}, nonequilibrium Green-function~\cite{PhysRevB78.235311,PhysRevLett.109.170402} and dynamical maps~\cite{PhysRevA71.020101,PhysRevA87.030101}.

There have been well-established theoretical investigations on the coherent dynamics
of open quantum systems coupled to environments which recover equilibrium instantly
from the interaction between the system and environments, generally assuming that the
environmental noise yields both stationary and Markovian statistical properties~\cite{PhysRevLett.88.228304,PhysRevLett.94.167002,PhysRevLett.96.097009,
RevModPhys.86.361,PhysRevA89.032114,JChemPhys.142.094106,*JChemPhys.142.094107}.
However, there are many situations where the nonequilibrium feature of the environments
plays an essential role in the coherent dynamics of an open quantum system.
The environment is no longer at thermal equilibrium which corresponds to physically that
the excited phonons of the environments are in certain nonstationary states initially.
For example, in some transient and ultrafast dynamical processes in physical or biological
systems which usually take place on a very short time scale, the initial nonstationary
states of the environmental excited phonons induced by the coupling to the quantum system
can not have the chance to recover equilibrium instantly~\cite{JChemPhys.133.241101,JPhysB45.154008,JChemPhys.139.024109,PhysRevA87.032338,PhysRevLett.112.246401,
PhysRevLett.115.257001,PhysRevA94.042110,JChemPhys.149.094107,SciRep.10.88,PhysRevA95.052104,EurophysLett.125.30007,NewJPhys.22.033039}.
On the other hand, a quantum system may not interact with a single environment where the coherent
dynamics of the system also depends strongly on the interaction between the sub-environments~\cite{PhysRevA90.042108,PhysRevA91.042111,PhysRevA92.012315}.
From this point of view, it is necessary to take the environmental nonequilibrium feature into
extensive consideration to study the coherent dynamics of an open quantum system with both
nonstationary and non-Markovian statistical properties of the environmental noise.

In this paper, we derive the generalized stochastic Liouville equation to investigate
the coherent dynamics of single molecule systems induced by environmental noise exhibiting
both nonstationary and non-Markovian features.
The generalized stochastic Liouville equation contains a generalized memory kernel
associated with both the intrinsic system Hamiltonian and the non-Markovian features
of the environmental noise.
In the limit case that the environment is in equilibrium and with no memory, the generalized
stochastic Liouville equation returns to the well-known framework established by Kubo.
In the presence of the nonstationary and non-Markovian RTN,
we express the exact analytical results of the coherence of the single
molecule systems. The analytical expressions of the system coherence are in good agreement with
that obtained by some other theoretical approaches.

\section{Generalized stochastic Liouville equation}

We consider a single molecule system coupled to a fluctuating environment
which displays both nonstationary and non-Markovian statistical properties.
Based on the spectral diffusion framework initiated by Kubo and Anderson,
the uncontrolled environmental degrees of freedom give rise to the stochastic
fluctuations in the energy eigenvalues of the system during the dynamical evolution as follows~\cite{JPhysSocJpn.9.316,JPhysSocJpn.9.935,RepProgPhys.80.056002}
\begin{equation}
\label{eq:Hami}
  H(t)=H_{0}+\delta H(t)=\hbar\sum_{n}[\omega_{n}+\epsilon_{n}(t)]|n\rangle\langle n|,
\end{equation}
where $H_{0}$ and $\delta H(t)$ are the intrinsic system Hamiltonian and its fluctuation term,
$|n\rangle$ is a set of orthonormal eigenstates, $\omega_{n}$ denotes the intrinsic frequency
of the state $|n\rangle$ and $\epsilon_{n}(t)$ is the environmental noise
which is subject to a stochastic process.

The dynamical evolution of the stochastic density matrix is governed by the Liouville equation
\begin{equation}
\label{eq:dynevo}
  \frac{\partial}{\partial t}\rho\bm(t;\delta H(t)\bm)=-\frac{i}{\hbar}[H(t),\rho\bm(t;\delta H(t)\bm)],
\end{equation}
where $\rho\bm(t;\delta H(t)\bm)$ is adopted to indicate that the system dynamics
depends on the stochastic fluctuation term $\delta H(t)$.
The dynamical evolution of the stochastic density matrix elements satisfies the differential equations
\begin{equation}
\label{eq:eleevo}
\begin{split}
\frac{\partial}{\partial t}\rho_{nn}\bm(t;\epsilon^{n}_{m}(t)\bm)&=0,\\
\frac{\partial}{\partial t}\rho_{nm}\bm(t;\epsilon^{n}_{m}(t)\bm)&=-i[\omega^{n}_{m}+\epsilon^{n}_{m}(t)]\rho_{nm}\bm(t;\epsilon^{n}_{m}(t)\bm),,
\end{split}
\end{equation}
where $\omega^{n}_{m}=\omega_{n}-\omega_{m}$ is the intrinsic frequency difference between states $|n\rangle$ and $|m\rangle$ and the off-diagonal element $\rho_{nm}\bm(t;\epsilon^{n}_{m}(t)\bm)$ depends on the composite environmental noise $\epsilon^{n}_{m}(t)=\epsilon_{n}(t)-\epsilon_{m}(t)$.
Due to the fact that $[H_{0},\delta H(t)]=0$, the quantum system undergoes pure decoherence
and the off-diagonal elements of the reduced density matrix change with time
while its populations are time independent and the diagonal elements of the reduced density matrix are constant.
The reduced density matrix can be, by averaging over different
realizations of the environmental noise, expressed as
\begin{equation}
\label{eq:reddenmat}
\begin{split}
 \rho(t)=&\langle\rho(t;\delta H(t))\rangle\\
        =&\sum_{n}\rho_{nn}(0)|n\rangle\langle n|+\sum_{\substack{n,m\\n\neq m}}\rho_{nm}(0)e^{-i\omega_{m}^{n}t}F_{m}^{n}(t)|n\rangle\langle m|,
\end{split}
\end{equation}
where $F_{m}^{n}(t)=\big\langle\exp[-i\int_{0}^{t}\epsilon^{n}_{m}(t')dt']\big\rangle$ is the decoherence function
describing the loss of superposition between states $|n\rangle$ and $|m\rangle$.
The coherence of the quantum system can be quantified in terms of the $l_{1}$ norm of coherence as~\cite{PhysRevLett.113.140401}
\begin{equation}
\label{eq:cohe}
  C_{l_{1}}(t)=\sum_{\substack{n,m\\n\neq m}}|\rho_{nm}(t)|=\sum_{\substack{n,m\\n\neq m}}|\langle\rho_{nm}\bm(t;\epsilon^{n}_{m}(t)\bm)\rangle|.
\end{equation}

In the following, we mainly focus on the derivation of the off-diagonal elements of the reduced density matrix by taking the average of the environmental noise.
We consider that the composite environmental noise $\epsilon^{n}_{m}(t)$ is subject to a stochastic
process which exhibits both nonstationary and non-Markovian features.
The non-Markovian feature of the noise process is characterized
by a generalized master equation for the conditional probability
which depends on its previous history
\begin{equation}
\label{eq:Mconpro}
  \frac{\partial}{\partial t}P(\varepsilon^{n}_{m},t|\varepsilon^{n'}_{m},t')=\int_{t'}^{t}K_{m}^{n}(t-\tau)\mathcal{M}_{\varepsilon^{n}_{m}}P(\varepsilon^{n}_{m},\tau|\varepsilon^{n'}_{m},t')d\tau,
\end{equation}
where the initial condition is given by $P(\varepsilon^{n}_{m},t'|\varepsilon^{n'}_{m},t')=\delta(\varepsilon^{n}_{m}-\varepsilon^{n'}_{m})$,
$K_{m}^{n}(t-\tau)$ denotes the memory kernel composite environmental noise $\epsilon^{n}_{m}(t)$ and
$\mathcal{M}_{\varepsilon^{n}_{m}}$ is a differential operator only involving derivatives
with respect to $\varepsilon^{n}_{m}$.
The composite environmental noise $\epsilon^{n}_{m}(t)$ is non-Markovian because the Chapman-Kolmogorov
equation is no longer valid unless it is memoryless, for instance, the memory kernel
is proportional to a $\delta$ function~\cite{Kampenbook}.
The nonstationary feature of the noise process $\epsilon^{n}_{m}(t)$ is characterized by the time
dependent single-time probability distribution~\cite{Kampenbook}
\begin{equation}
\label{eq:onepro}
  P(\varepsilon^{n}_{m},t)=\int P(\varepsilon^{n}_{m},t|\varepsilon^{n}_{m0},0)P(\varepsilon^{n}_{m0},0)d\varepsilon^{n}_{m0},
\end{equation}
where $P(\varepsilon^{n}_{m0},0)$ is a nonstationary probability which
represents that the environmental states are nonstationary initially.
In contrast to the usual treatment, the statistical properties of the
environmental noise are nonstationary, corresponding physically to
impulsively excited phonons of the environment with sharply defined
initial values at $t=0$. Under this assumptions, the environment
is not at thermal equilibrium.
For the case $P(\varepsilon^{n}_{m0},0)=\lim\limits_{t\rightarrow\infty} P(\varepsilon^{n}_{m},t|\varepsilon^{n}_{m0},0)$,
the environmental noise $\epsilon^{n}_{m}(t)$ is stationary and the environment is at thermal equilibrium~\cite{Gardinerbook1,Kampenbook}.

To generalize the stochastic Liouville equation to the coherent dynamics in the
presence of environments exhibiting both nonstationary and non-Markovian features,
we first introduce the bivariate process $\{\varrho_{nm},\varepsilon^{n}_{m}\}$ and define the joint
probability~\cite{Kampenbook,Riskenbook}
\begin{equation}
\label{joipro}
\begin{split}
  P(\varrho_{nm},\varepsilon^{n}_{m},t)&=P(\varrho_{nm},t|\varepsilon^{n}_{m},t)P(\varepsilon^{n}_{m},t)\\
                            &=\left\langle\delta\bm(\rho_{nm}(t)-\varrho_{nm}\bm)\delta\bm(\epsilon^{n}_{m}(t)-\varepsilon^{n}_{m}\bm)\right\rangle.
\end{split}
\end{equation}
Consequently, the time evolution of the joint probability $P(\varrho_{nm},\varepsilon^{n}_{m},t)$ describes a flow in $(\varrho_{nm},\varepsilon^{n}_{m})$ space
\begin{equation}
\label{flow}
\begin{split}
\frac{\partial}{\partial t}P(\varrho_{nm},\varepsilon^{n}_{m},t)=&i(\omega^{n}_{m}+\varepsilon^{n}_{m})\frac{\partial}{\partial \varrho_{nm}}\varrho_{nm}P(\varrho_{nm},\varepsilon^{n}_{m},t)\\
 &+\int_{0}^{t}K_{m}^{n}(t-t')\mathcal{M}_{\varepsilon^{n}_{m}}P(\varrho_{nm},\varepsilon^{n}_{m},t')dt',
\end{split}
\end{equation}
where the initial condition satisfies $P(\varrho_{nm},\varepsilon^{n}_{m},0)=\delta\bm(\rho_{nm}(0)-\varrho_{nm0}\bm)P(\varepsilon^{n}_{m0},0)$ and it has been assumed that the environmental noise $\epsilon^{n}_{m}(t)$ is not influenced by the system and that there is no initial correlation between the system and its environment.

The off-diagonal element of the reduced density matrix of the quantum system can be
obtained in $(\varrho_{nm},\varepsilon^{n}_{m})$ space by means of two separated averages.
We first define the partial average over $\varrho_{nm}$ for fixed $\varepsilon^{n}_{m}$ as
\begin{equation}
\label{parave}
  \rho_{nm}(\varepsilon^{n}_{m},t)=\int\varrho_{nm}P(\varrho_{nm},\varepsilon^{n}_{m},t)d\varrho_{nm}.
\end{equation}
By multiplying Eq.~(\ref{flow}) with $\varrho_{nm}$ and integrating, we obtain the generalized stochastic Liouville equation
\begin{equation}
\label{gsLe}
\begin{split}
  \frac{\partial}{\partial t}\rho_{nm}(\varepsilon^{n}_{m},t)=&-i(\omega^{n}_{m}+\varepsilon^{n}_{m})\rho_{nm}(\varepsilon^{n}_{m},t)\\
  &+\int_{0}^{t}\mathcal{K}_{m}^{n}(t-t')\mathcal{M}_{\varepsilon^{n}_{m}}\rho_{nm}(\varepsilon^{n}_{m},t')dt',
\end{split}
\end{equation}
where $\mathcal{K}_{m}^{n}(t-t')=K_{m}^{n}(t-t')e^{-i\omega^{n}_{m}(t-t')}$ denotes the generalized memory kernel which is associated with both the intrinsic frequency difference $\omega^{n}_{m}$ between states $|n\rangle$ and $|m\rangle$ and the memory kernel of the composite environmental noise $\epsilon^{n}_{m}(t)$. Here the generalized stochastic Liouville equation~\eqref{gsLe} contains both the unitary and nonunitary parts of the evolution arising from the environmental nonstationary and non-Markovian features, respectively.
Consequently, the off-diagonal element of the reduced density matrix $\rho_{nm}(t)$ can be obtained by the complete average
\begin{equation}
\label{eq:comave}
  \rho_{nm}(t)=\int \rho_{nm}(\varepsilon^{n}_{m},t)d\varepsilon^{n}_{m}.
\end{equation}
It is worth mentioning that when the composite environmental noise is memoryless, i.e. $K_{m}^{n}(t-t')=\delta(t-t')$, the generalized stochastic Liouville equation returns to the well-known framework established by Kubo~\cite{JMathPhys.4.174}.
It is convenient to derive the reduced density matrix of the quantum system by means of the generalized stochastic Liouville equation since we just need to know the statistical properties of the environmental noise, e.g., stationary or nonstationary, and Markovian or non-Markovian, instead of its statistical characteristics, such as the correlation function of each order.
However, the differential operator $\mathcal{M}_{\varepsilon^{n}_{m}}$ describing the statistical properties of the environmental noise in Eq.~\eqref{eq:Mconpro} is of arbitrary form or even nonlinear.
Therefore, the reduced density matrix can be analytically solved only for a few special cases but we can obtain its numerical solution by means of the generalized stochastic Liouville equation.

\section{Theoretical results and discussion}

In this section, we study a special and important case of the coherent dynamics in the presence of the environments
exhibiting nonstationary and non-Markovian RTN features.
We will derive analytically the off-diagonal element of the reduced density matrix of the the quantum system by means of the generalized stochastic Liouville equation.
We assume that the composite environmental noise $\epsilon^{n}_{m}(t)$ obeys a nonstationary
non-Markovian RTN process associated with an initially nonstationary distribution
and an environmental memory kernel, which jumps randomly between the values $\pm\nu_{m}^{n}$
with the switching rate $\lambda_{m}^{n}$~\cite{PhysRevA94.042110,JChemPhys.149.094107,SciRep.10.88}.
The time evolution of the conditional probability of the composite environmental noise
$\epsilon^{n}_{m}(t)$ is governed by the generalized master equation~\cite{PhysRevE50.2668}
\begin{equation}
\label{equ31}
\begin{split}
  \frac{\partial}{\partial t}P(+\nu_{m}^{n},t|\varepsilon^{n'}_{m},t')=&\int_{t'}^{t} \lambda_{m}^{n}K_{m}^{n}(t-\tau)\big[P(-\nu_{m}^{n},\tau|\varepsilon^{n'}_{m},t')\\
                                              &-P(+\nu_{m}^{n},\tau|\varepsilon^{n'}_{m},t')\big]d\tau,\\
  \frac{\partial}{\partial t}P(-\nu_{m}^{n},t|\varepsilon^{n'}_{m},t')=&\int_{t'}^{t}
  \lambda_{m}^{n}K_{m}^{n}(t-\tau)\big[ P(+\nu_{m}^{n},\tau|\varepsilon^{n'}_{m},t')\\
                                              &-P(-\nu_{m}^{n},\tau|\varepsilon^{n'}_{m},t')\big]d\tau.
\end{split}
\end{equation}
The environment is not at thermal equilibrium, corresponding to the
initially nonstationary distribution for environmental noise as
\begin{equation}
\label{equ32}
  P(\varepsilon^{n}_{m0},0)=\frac{1}{2}(1-a_{m}^{n})\delta_{\varepsilon^{n}_{m0},-\nu_{m}^{n}}+\frac{1}{2}(1+a_{m}^{n})\delta_{\varepsilon^{n}_{m0},+\nu_{m}^{n}},
\end{equation}
where $\varepsilon^{n}_{m0}=\pm\nu_{m}^{n}$ and $a_{m}^{n}$ is the nonequilibrium parameter and $-1\leq a_{m}^{n}\leq1$.
For the case $a_{m}^{n}=0$, the environmental states are stationary initially
corresponding to that the environment is in equilibrium.

Based on Eq.~(\ref{gsLe}), the generalized stochastic Liouville equation for the
coherent dynamics induced by nonstationary and non-Markovian RTN can be expressed as
\begin{equation}
\label{gsLeRT}
\begin{split}
\frac{\partial}{\partial t}\rho_{nm}(+\nu_{m}^{n},t)=&-i(\omega_{m}^{n}+\nu_{m}^{n})\rho_{nm}(+\nu_{m}^{n},t)+\int_{0}^{t}\lambda_{m}^{n}\\
&\times K_{m}^{n}(t-t')e^{-i\omega_{m}^{n}(t-t')}\big[\rho_{nm}(-\nu_{m}^{n},t')\\
&-\rho_{nm}(+\nu_{m}^{n},t')\big]dt',\\
\frac{\partial}{\partial t}\rho_{nm}(-\nu_{m}^{n},t)=&-i(\omega_{m}^{n}-\nu_{m}^{n})\rho_{nm}(-\nu_{m}^{n},t)+\int_{0}^{t}\lambda_{m}^{n}\\
&\times K_{m}^{n}(t-t')e^{-i\omega_{m}^{n}(t-t')}\big[\rho_{nm}(+\nu_{m}^{n},t')\\
&-\rho_{nm}(-\nu_{m}^{n},t')\big]dt',
\end{split}
\end{equation}
where the initial conditions are given by $\rho_{nm}(\pm\nu_{m}^{n},0)=\frac{1}{2}\big(1\pm a_{m}^{n}\big)\rho_{nm}(0)$.
By means of Laplace transform, the off-diagonal element of the reduced density matrix $\rho_{nm}(t)$ can be written in a sum as
\begin{equation}
\label{comaveRT}
\begin{split}
 \rho_{nm}(t)&=\rho_{nm}(+\nu_{m}^{n},t)+\rho_{nm}(-\nu_{m}^{n},t)\\
             &=e^{-i\omega_{m}^{n}t}F_{m}^{n}(t)\rho_{nm}(0),
 \end{split}
\end{equation}
where the decoherence function $F_{m}^{n}(t)$ can be analytically solved as
\begin{equation}
\label{eq:depfacexp}
\begin{split}
   F_{m}^{n}(t)&= \mathcal{L}^{-1}[\widetilde{F}_{m}^{n}(p)],\\
    \widetilde{F}_{m}^{n}(p) &= \frac{p+2\lambda_{m}^{n}\widetilde{K}_{m}^{n}(p)+ia_{m}^{n}\nu_{m}^{n}}{p\left[p+2\lambda_{m}^{n}\widetilde{K}_{m}^{n}(p)\right]+(\nu_{m}^{n})^{2}}.
\end{split}
\end{equation}
Here $\mathscr{L}^{-1}$ indicates the inverse Laplace transform, $\widetilde{K}_{m}^{n}(p)$ is the Laplace transform of the memory kernel and the initial condition is given by $F_{m}^{n}(0)=1$.
Clearly, it is more convenient to calculate the decoherence function induced by a nonstationary and non-Markovian RTN process based on the expression derived in Eq.~\eqref{eq:depfacexp} rather than to derive a closed equation for the characteristic function based on the statistical characteristics of the environmental noise as we did in Refs.~\cite{PhysRevA94.042110,JChemPhys.149.094107}.

%Based on the expression of the off-diagonal element of the reduced density matrix $\rho_{nm}(t)$ in Eq.~\eqref{comaveRT}, the dynamical evolution of the reduced density matrix of the system is described by a time-local quantum equation
%\begin{equation}
%\label{eq:dynevo}
%  \frac{\partial}{\partial t}\rho(t)=-\frac{i}{\hbar}[H_{0}+H_{\mathrm{LS}}(t),\rho(t)]+\sum_{\substack{n,m\\n\neq m}}\gamma_{m}^{n}(t)[\Sigma_{m}^{n},[\Sigma_{m}^{n},\rho(t)]],
%\end{equation}
%where the operator $\Sigma_{m}^{n}=|n\rangle\langle n|-|m\rangle\langle m|$,

\subsubsection{Composite memory kernel}

We first consider the case that the composite environmental noise $\epsilon^{n}_{m}(t)$ is with a composite form of memory kernel
\begin{equation}
\label{eq:memker1}
 K_{m}^{n}(t-\tau)=w_{m}^{n}\delta(t-\tau)+(1-w_{m}^{n})\kappa_{m}^{n}e^{-\kappa_{m}^{n}(t-\tau)},
\end{equation}
where $0\leq w_{m}^{n}\leq1$ denotes a weight factor for Markovian and non-Markovian features of the environmental noise and $\kappa_{m}^{n}$ is the memory decay rate of the exponential kernel.
In Laplace domain, the composite memory kernel yields $\widetilde{K}_{m}^{n}(p)=w_{m}^{n}+(1-w_{m}^{n})\frac{\kappa_{m}^{n}}{p+\kappa_{m}^{n}}$ and the decoherence function in Eq.~\eqref{eq:depfacexp} can be expressed as
\begin{equation}
\label{eq:depfacexp1}
\begin{split}
   F_{m}^{n}(t)&= \mathcal{L}^{-1}[\widetilde{F}_{m}^{n}(p)],\\
    \widetilde{F}_{m}^{n}(p) &= \frac{p+2\lambda_{m}^{n}\big[w_{m}^{n}+(1-w_{m}^{n})\frac{\kappa_{m}^{n}}{p+\kappa_{m}^{n}}\big]+ia_{m}^{n}\nu_{m}^{n}}
    {p\left\{p+2\lambda_{m}^{n}\big[w_{m}^{n}+(1-w_{m}^{n})\frac{\kappa_{m}^{n}}{p+\kappa_{m}^{n}}\big]\right\}+(\nu_{m}^{n})^{2}}.
\end{split}
\end{equation}

In the memoryless limit of the environmental noise, namely, the weight $w_{m}^{n}=1$ or the decay rate of the exponential memory kernel $\kappa_{m}^{n}\rightarrow+\infty$.
In this case, the Laplace domain decoherence function in Eq.~\eqref{eq:depfacexp1} can be reduced to
\begin{equation}
\label{eq:FlRTN1}
 \widetilde{F}_{m}^{n}(p)=\frac{p+2\lambda_{m}^{n}+ia_{m}^{n}\nu_{m}^{n}}{p(p+2\lambda_{m}^{n})+(\nu_{m}^{n})^{2}}.
\end{equation}
By taking the inverse Laplace transform of Eq.~(\ref{eq:FlRTN1}), we can express the decoherence function in time domain as
\begin{equation}
\label{eq:FtRTN}
F_{m}^{n}(t)= e^{-\lambda_{m}^{n}t}
\begin{cases}
\begin{split}
&\Big[\cosh(\beta_{m}^{n}t)+\frac{\lambda_{m}^{n}}{\beta_{m}^{n}}\sinh(\beta_{m}^{n}t)\Big] \\
&+i\frac{a_{m}^{n}\nu_{m}^{n}}{\beta_{m}^{n}}\sinh(\beta_{m}^{n}t),\\ %& \nu_{m}^{n}<\lambda_{m}^{n},\\
&\big(1+\lambda_{m}^{n}t\big)+ia_{m}^{n}\lambda_{m}^{n}t,\\ %& \nu_{m}^{n}=\lambda_{m}^{n},\\
&\Big[\cos(\beta_{m}^{n}t)+\frac{\lambda_{m}^{n}}{\beta_{m}^{n}}\sin(\beta_{m}^{n}t)\Big]\\
&+i\frac{a_{m}^{n}\nu_{m}^{n}}{\beta_{m}^{n}}\sin(\beta_{m}^{n}t), %& \nu_{m}^{n}>\lambda_{m}^{n},
\end{split}
\end{cases}
\end{equation}
with $\beta_{m}^{n}=\sqrt{|(\lambda_{m}^{n})^{2}-(\nu_{m}^{n})^{2}|}$ for $\nu_{m}^{n}<\lambda_{m}^{n}$, $\nu_{m}^{n}=\lambda_{m}^{n}$
and $\nu_{m}^{n}>\lambda_{m}^{n}$, respectively.
This expression of the decoherence function in Eq.~\eqref{eq:FtRTN} is consistent with that obtained in Ref.~\cite{JChemPhys.149.094107} and it returns to the well-known results~\cite{PhysRevB78.201302,PhysStatusSolidiB246.1018,PhysRevA89.012330} when the environment is in equilibrium, namely, the statistical property of the environmental noise is stationary.

\subsubsection{Modulatable memory kernel}

We now consider the case that the composite environmental noise $\epsilon^{n}_{m}(t)$ is subject to a modulatable non-Markovian process with the memory kernel
\begin{equation}
\label{eq:memker2}
 K_{m}^{n}(t-\tau)=\kappa_{m}^{n}e^{-\kappa_{m}^{n}(t-\tau)}\cos[\Omega_{m}^{n}(t-t')],
\end{equation}
where $\Omega_{m}^{n}$ denotes the external modulation frequency of the environment~\cite{PhysRevA30.568,PhysRevE83.041104}.
The modulatable memory kernel becomes an exponential form when there is no environmental frequency modulation $\Omega_{m}^{n}=0$.
The memory kernel in Laplace domain yields $\widetilde{K}_{m}^{n}(p)=\frac{
\kappa_{m}^{n}(p+\kappa_{m}^{n})}{(p+\kappa_{m}^{n})^{2}+(\Omega_{m}^{n})^{2}}$
and the decoherence function in Eq.~\eqref{eq:depfacexp} can be written as
\begin{equation}
\label{eq:depfacexp2}
\begin{split}
   F_{m}^{n}(t)&= \mathcal{L}^{-1}[\widetilde{F}_{m}^{n}(p)],\\
 \widetilde{F}_{m}^{n}(p) &=\frac{p+\frac{2\lambda_{m}^{n}\kappa_{m}^{n}(p+\kappa_{m}^{n})}{(p+\kappa_{m}^{n})^{2}+(\Omega_{m}^{n})^{2}}+ia_{m}^{n}\nu_{m}^{n}}
 {p\left[p+\frac{2\lambda_{m}^{n}\kappa_{m}^{n}(p+\kappa_{m}^{n})}{(p+\kappa_{m}^{n})^{2}+(\Omega_{m}^{n})^{2}}\right]+(\nu_{m}^{n})^{2}}.
\end{split}
\end{equation}
The expression of the decoherence function in Eq.~\eqref{eq:depfacexp2} returns to the result obtained in Ref.~\cite{SciRep.10.88} in the case that the statistical property of the
environmental noise is stationary, namely, when the environment is in equilibrium.
For the case of no environmental frequency modulation $\Omega_{m}^{n}=0$,
the memory kernel in Eq.~\eqref{eq:memker2} becomes an exponential form
and the decoherence function in Eq.~\eqref{eq:depfacexp2} is in consistent with
that derived in Refs.~\cite{PhysRevA94.042110,JChemPhys.149.094107}.

\section{Conclusions}

We have proposed the generalized stochastic Liouville equations to investigate
the coherent dynamics of single molecule systems coupled to the environment of
which the statistical properties are both nonstationary and non-Markovian.
The coherence of the quantum system can be obtained by means of the generalized
stochastic Liouville equation in two separated averages.
The generalized stochastic Liouville equation yields a generalized master equation
containing a memory kernel related to the intrinsic system Hamiltonian and the
non-Markovian features of the environmental noise
and it returns to the well-known framework established by Kubo in the limit case
that the environment is in equilibrium and memoryless.
We express the exact analytical results of the coherent dynamics of the single
molecule systems induced by the nonstationary and non-Markovian RTN.
The analytical expressions of the system coherence are in well consistent with
that obtained by means of some other theoretical approaches.

%Acknowledgments of the paper
\begin{acknowledgments}
This work is supported by the National Basic Research Program of China (Grant No. 2015CB921004) and the National Natural Science Foundation of China (Grant Nos. 11674196 and 11947033).
X.C. also acknowledges the support from the Doctoral Research Fund of Shandong Jianzhu University (Grant No. XNBS1852).
\end{acknowledgments}

%\end{CJK*}

%References
%\bibliography{References}
%

\end{document}